\documentclass[3p,times,twocolumn]{elsarticle}

\usepackage{ecrc}

\volume{00}

\firstpage{1}

\journalname{Nuclear Physics B Proceedings Supplement}

\runauth{A.~Francis et al.}

\jid{nuphbp}

\jnltitlelogo{Nuclear Physics B Proceedings Supplement}

\usepackage{amssymb}

\usepackage{amsmath}
\usepackage[tight,TABTOPCAP]{subfigure}
\usepackage{graphicx}
\usepackage{grffile}
\usepackage{relsize}

\biboptions{sort&compress}

\usepackage[figuresright]{rotating}

\newcommand{\psibar}{\overline \psi}

\graphicspath{{figures/}}

\begin{document}

\begin{frontmatter}



  \dochead{}

  \title{Lattice QCD Studies of the Leading Order Hadronic Contribution
    to the \\ Muon $g-2$}

  \author[prisma]{Anthony Francis}
  \author[prisma,him]{Vera G\"ulpers}
  \author[ift]{Gregorio Herdo\'iza}
  \author[prisma]{Georg von Hippel}
  \author[prisma]{Hanno Horch}
  \author[swansea]{Benjamin J\"{a}ger}
  \author[prisma,him]{Harvey B. Meyer}
  \author[prisma]{Eigo Shintani}
  \author[prisma,him]{Hartmut Wittig}

  \address[prisma]{PRISMA Cluster of Excellence, Institut f{\"u}r Kernphysik,\\
    Johannes Gutenberg-Universit{\"a}t, 55099 Mainz, Germany}
  \address[him]{Helmholtz Institute Mainz,\\ Johannes
    Gutenberg-Universit{\"a}t, 55099 Mainz, Germany}
  \address[ift]{Instituto de F\'isica Te\'orica UAM/CSIC and
    Departamento de F\'isica Te\'orica,\\
    Universidad Aut\'onoma de Madrid, Cantoblanco, E-28049
    Madrid, Spain}
  \address[swansea]{Department of Physics, Swansea University, Swansea, United Kingdom}

  \begin{abstract}

    The anomalous magnetic moment of the muon, $g_\mu-2$, is one of
    the most promising observables to identify hints for physics
    beyond the Standard Model. QCD contributions are currently
    responsible for the largest fraction of the overall theoretical
    uncertainty in $g_\mu-2$. The possibility to determine these
    hadronic contributions from first principles through lattice QCD
    calculations has triggered a number of recent studies.  Recent
    proposals to improve the accuracy of lattice determinations are
    reported. We present an update of our studies of the leading-order
    hadronic contribution to $g_\mu-2$ with improved Wilson fermions.

  \end{abstract}


\end{frontmatter}

%%\linenumbers


\section{Introduction}
\label{sec:intro}

%%%%%%%%%%%%%%%%%%%%%%%%%%%%%%%%%%%%%%%%%%%%%%%%%%%%%%%%%%%%%%%%%%%

The anomalous magnetic moment of the muon, $a_\mu=(g_\mu-2)/2$, is a
remarkable example of a quantity that can be studied with very high
accuracy on both the experimental and the theoretical sides. The 0.5
ppm uncertainty of the current experimental value allows to probe
contributions from electromagnetic, strong and weak interactions. The
Standard Model (SM) prediction has reached a comparable
precision\,~\cite{Agashe:2014kda},
\begin{align}
  \label{eq:gm2res}
  \hspace*{-1.4cm}
  a_\mu^{\rm th} &= 116\,591\,803\,{(42)}\,(26)\,(01)~\cdot 10^{-11}~\,{[0.4\,{\rm ppm}]}\,,\nonumber\\
  a_\mu^{\rm exp} &= 116\,592\,091\,(54)\,(33)~\cdot 10^{-11} ~~~\,{[0.5\,{\rm ppm}]}\,.
\end{align}
The deviation between theory and experiment currently amounts to a 3.6
$\sigma$ effect. The next generation of experiments at
Fermilab~\cite{Carey:2009zzb} and J-PARC~\cite{Benayoun:2014tra} aims
at a reduction of the uncertainty in $a_\mu$ by a factor of four. Such
a precision will substantially enhance the sensitivity to physics
beyond the SM. It is, however, equally important to examine the
reliability of the current SM prediction and to attempt to reduce its
uncertainty to the level of the forthcoming experimental results. The
SM result has recently profited from an outstanding achievement in
determining the QED contribution up to 5-loop
order~\cite{Aoyama:2012wk}. Theory errors in eq.~(\ref{eq:gm2res})
arise from lowest-order hadronic (HLO), higher-order hadronic and
electroweak contributions, respectively. The SM error is thus markedly
dominated by QCD dynamics and, in particular, by HLO vacuum
polarisation effects.

The HLO vacuum polarisation contribution, $a_\mu^{\rm HLO}$, can be
obtained by a dispersive approach that combines basic properties of
the theory -- such as analyticity and unitarity -- with experimental
input. A collection of recent
measurements~\cite{Davier:2010nc,Hagiwara:2011af,Benayoun:2012wc} of
inclusive hadronic cross-sections, $\sigma(e^+e^- \to {\rm hadrons})$,
has allowed to reach a $0.6\%$ precision on the LO hadronic
contribution, $a_\mu^{\rm HLO} = 6923(42) \cdot
10^{-11}$~\cite{Agashe:2014kda}. A persistent $\sim 3\sigma$ deviation
between the analyses of the $\pi^+\pi^-$ channel by BaBar and KLOE has
an impact on the SM prediction. This is being investigated by several
experiments~\cite{Benayoun:2014tra}. Conversely, the tension in the
results for $a_\mu^{\rm HLO}$ based on $e^+e^-$ and $\tau$ data has
recently been reduced below the 2\,$\sigma$
level~\cite{Jegerlehner:2011ti,Benayoun:2012wc}.

Since the dispersion relation results largely depend on experimental
data, it is desirable to consider also an independent approach based
on first principles. A determination of $a_\mu^{\rm HLO}$ along these
lines can be achieved through lattice QCD. A number of
studies~\cite{Blum:2002ii,Gockeler:2003cw,Aubin:2006xv,Feng:2011zk,DellaMorte:2011aa,Boyle:2011hu,Burger:2013jya}
have demonstrated the potential of this approach. It is nonetheless
still a considerable challenge for the lattice studies to reach the
sub-percent accuracy of the dispersion relation result. There has
recently been an intense activity in order to device new ways of
improving the accuracy of the lattice determinations of $a_\mu^{\rm
  HLO}$.

%%%%%%%%%%%%%%%%%%%%%%%%%%%%%%%%%%%%%%%%%%%%%%%%%%%%%%%%%%%%%%%%%%%

\section{Lattice QCD Determinations of $a_\mu^{\rm HLO}$}
\label{sec:latamu}

\begin{figure}[t!]
  \centering
  \includegraphics[scale=0.69]{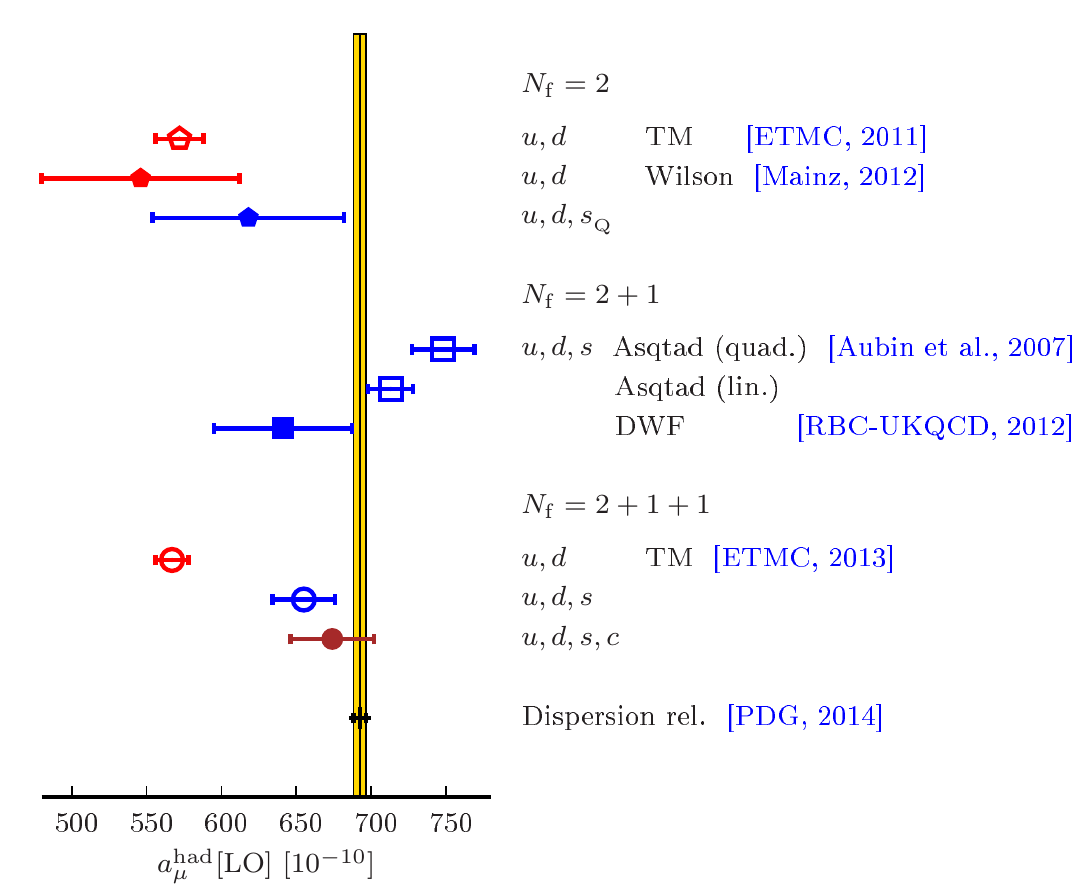}
  \caption{Comparison of lattice determinations of $a_\mu^{\rm
      HLO}$~\cite{Feng:2011zk,DellaMorte:2011aa,Aubin:2006xv,Boyle:2011hu,Burger:2013jya}. The
    number of flavours in the sea is labelled by $N_{\rm f}$ while the
    flavour content in the valence sector, appearing in
    eq.~(\ref{eq:current}), is denoted by $u$,~$d$,~$s$ and $c$. The
    dispersion relation approach -- with a $0.6\%$ relative
    precision~\cite{Agashe:2014kda} -- is the denoted by the yellow
    vertical band.}
  \label{fig:gm2comp}
\end{figure}

The hadronic vacuum polarisation tensor, depending on the Euclidean
momentum $Q$, reads
\begin{equation}\label{eq:pol_tensor}
  \Pi_{\mu\nu}(Q)= \int d^4x
  \,e^{iQx} \,\langle J_\mu(x)J_\nu(0) \rangle \,,
\end{equation}
where the flavour singlet vector current is given by,
\begin{equation}
  J_\mu(x)=\sum_{{\rm f}=u,\,d,\,s,\,c,\dots} \, Q_{\rm f} \, \psibar_{\rm
    f}(x)\gamma_\mu \psi_{\rm f}(x) \,.
  \label{eq:current}
\end{equation}
$Q_{\rm f}$ is the electric charge of the quark flavour ${\rm
  f}$. Euclidean invariance and current conservation
imply,
\begin{equation}
  \Pi_{\mu\nu}(Q)=(Q_\mu Q_\nu - \delta_{\mu\nu} Q^2)
  \,\Pi(Q^2)\,.
  \label{eq:vpcont}
\end{equation}
The VPF $\Pi(Q^2)$ can be decomposed into non-singlet and singlet
contributions. The subtracted VPF, $\widehat{\Pi}(Q^2) =
\Pi(Q^2)-\Pi(0)$, is free of ultraviolet divergences and can be
convoluted with a known analytic  kernel function $K(Q^2,m_\mu)$ to derive
the {\it standard representation} for $a_\mu^{\rm
  HLO}$~\cite{deRafael:1993za,Blum:2002ii} currently being used on the
lattice,
\begin{equation}
  a_\mu^{\rm HLO} = 4\alpha^2 \,
  \int_{0}^{\infty} dQ^2 \, K(Q^2,m_\mu) \,\widehat{\Pi}(Q^2)\,,
  \label{eq:amulat}
\end{equation}
where $m_\mu$ is the muon mass. A comparison of lattice QCD
determinations of $a_\mu^{\rm
  HLO}$~\cite{Feng:2011zk,DellaMorte:2011aa,Aubin:2006xv,Boyle:2011hu,Burger:2013jya}
is shown in Fig.~\ref{fig:gm2comp}.
The present uncertainty from the lattice computations is larger than
the $0.6\%$ precision of the dispersion relation approach. With the
current accuracy, it is still challenging to isolate the relative
contributions from dynamical strange and charm quarks. However, the
$s$ and $c$ valence contributions -- while being significantly smaller
than those from $u,d$ quark flavours -- can be determined with
relatively good precision~\cite{Burger:2013jya,Chakraborty:2014mwa}.

The uncertainties of the lattice QCD results of $a_\mu^{\rm HLO}$ have
multiple origins. In the next section we outline the main sources of
errors affecting these computations and report about the recent
proposals to address them.

%%%%%%%%%%%%%%%%%%%%%%%%%%%%%%%%%%%%%%%%%%%%%%%%%%%%%%%%%%%%%%%%%%%

\section{Behaviour of the VPF at Low $Q^2$}
\label{sec:lowq2}

\begin{figure}[t!]
  \centering
  \includegraphics[scale=0.69]{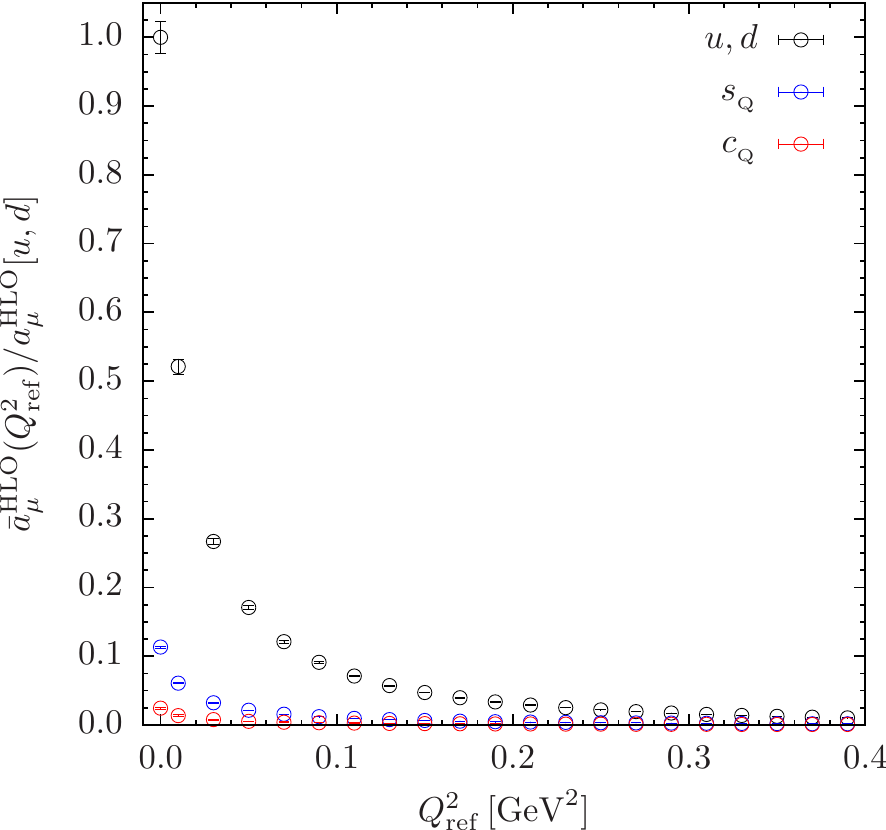}
  \caption{Momentum dependence of $\bar{a}_\mu^{\rm HLO}(Q^2_{\rm
      ref})$, defined in eq.~(\ref{eq:amubar}), and coinciding with
    $a_\mu^{\rm HLO}$ at $Q^2_{\rm ref}=0$. The y-axis is normalised
    by $a_\mu^{\rm HLO}[u,d]$. The region, $Q^2_{\rm ref} \gtrsim
    0.4\,{\rm GeV}^2$, is observed to contribute very little to
    $a_\mu^{\rm HLO}$. When increasing the quark mass from the
    mass-degenerate $u,d$ quark sector to the strange and charm
    regions, a strong suppression of the contribution to $a_\mu^{\rm
      HLO}$ is observed. The current accuracy however requires these
    various contributions to be included.}
  \label{fig:amubar}
\end{figure}

A crucial aspect of the computation of $a_\mu^{\rm HLO}$ is to
constrain with accurate lattice data the $Q^2$ region where the
integrand in eq.~(\ref{eq:amulat}) is large. In practice, this region
is in the neighbourhood of $Q^2 \approx m_\mu^2/4 \approx 0.003\,{\rm
  GeV}^2$. However, this low-$Q^2$ regime poses serious problems for
lattice studies based on an evaluation of $\Pi(Q^2)$ from
eq.~(\ref{eq:vpcont}), since the transverse projector on the r.h.s
vanishes at $Q^2=0$. In finite volume with periodic boundary
conditions, the minimal momentum is quantised in units of the lattice
size $L$, by $Q^2_{\rm min} = (2\pi/L)^2$. Directly probing the
dominant region, $Q^2 \approx m_\mu^2/4$, would require values of $L
\approx 20$\,fm that are far beyond what is achievable with
present-day resources. Furthermore, in this small $Q^2$ regime,
long-distance QCD effects induce large fluctuations on the VPF. To
illustrate these effects an auxiliary observable,
$\bar{a}_\mu^{\rm had}(Q^2_{\rm ref})$, is defined as follows,
\begin{equation}
  \hspace*{-0.45cm}
  \bar{a}_\mu^{\rm HLO}(Q^2_{\rm ref}) = 4\alpha^2 \int_{Q^2_{\rm ref}}^{\infty}
  dQ^2 K(Q^2) \left[\Pi(Q^2) - {\Pi(Q^2_{\rm
        ref})}\right].
  \label{eq:amubar}
\end{equation}
This quantity coincides with $a_\mu^{\rm HLO}$ in the limit $Q^2_{\rm
  ref} \to 0$. Fig.~\ref{fig:amubar} shows the dependence of
$\bar{a}_\mu^{\rm HLO}$ on $Q^2_{\rm ref}$. The $u,d$ valence quark
contribution is dominated by the region $Q^2_{\rm ref} \lesssim
0.4\,{\rm GeV}^2$. Moreover, the relative error on $\bar{a}_\mu^{\rm
  HLO}$ increases when reducing $Q^2_{\rm ref}$. A clear hierarchy is
observed in the size of the $(u, d)$, $s$ and $c$ valence
contributions. In spite of that, the current accuracy is at a level
that renders the inclusion of these effects appropriate.
A number of ideas have been recently put forward to tackle the issue
of reaching the small $Q^2$ regime.

\paragraph{Partially Twisted Boundary Conditions}

To circumvent the limitation of having access only to a restricted set
of low momentum values, periodic boundary conditions for the valence
quark fields can be replaced by twisted boundary
conditions~\cite{Bedaque:2004kc,deDivitiis:2004kq,Sachrajda:2004mi}. A
denser set of momenta can thus be reached~\cite{DellaMorte:2011aa} in
a region closer to $Q^2 \approx m_\mu^2$, at the price of additional
numerical effort and of small systematic effects from the breaking of
isospin
symmetry~\cite{Aubin:2013daa,Horch:2013lla,Gregory:2013taa}. The
increasing fluctuations in $\Pi(Q^2)$ at small $Q^2$ values are however
still present when adopting this procedure.

\paragraph{Extrapolation to $Q^2=0$}

The integrand in eq.~(\ref{eq:amulat}) is peaked at low $Q^2$ where
lattice data are not available. Moreover, an extrapolation of
$\Pi(Q^2)$ to $Q^2 \to 0$ is needed when relying on the {\it standard
  representation} for $a_\mu^{\rm HLO}$. The estimate of the
systematic effects associated with this extrapolation is one of the
crucial aspects of present lattice calculations.  Parametrisations of
the $Q^2$ dependence of the VPF based on vector meson dominance can
introduce a model dependence that is difficult to
quantify. Alternatively, Pad\'e approximants supply a
model-independent and systematically improvable description of the
$Q^2$ behaviour of
$\Pi(Q^2)$~\cite{Aubin:2012me,Golterman:2013vca,DellaMorte:2011aa}.
Correlations among $Q^2$ data points and the increasing number of fit
parameters limits the order of the Pad\'e approximants that can be
reached for the purpose of testing the convergence properties of the
series. This problem can be alleviated by restricting the use of
Pad\'e fits to the low momentum region, $Q^2 \lesssim 0.4\,{\rm
  GeV}^2$, which is known to provide the bulk on the contribution to
$a_\mu^{\rm HLO}$, see Fig.~\ref{fig:amubar}. By splitting the bounds
of the integral in eq.~(\ref{eq:amulat}) into low and high $Q^2$
intervals, a dedicated analysis of each of these regions can lead to
an additional handle on the assessment of systematic
effects~\cite{Golterman:2014ksa,Golterman:2014wfa}.

\paragraph{Momentum Derivatives of the Vacuum Polarisation}

A complementary way to scrutinise the difficulties encountered in
the low-$Q^2$ region is to consider derivatives with respect to
momentum of the vacuum polarisation.

By applying derivatives of the vacuum polarisation tensor in
eq.~(\ref{eq:vpcont}) with respect to $Q_\mu$ and $Q_\nu$, it is
possible to extract $\Pi(Q^2)$ and, in particular, to isolate
$\Pi(0)$. These derivatives have formally been applied in order to
rewrite $\Pi(0)$ in terms of suitable correlation functions involving
the integrated insertion of currents~\cite{deDivitiis:2012vs}. The
availability of $\Pi(0)$ then allows to reach the dominant momentum
region, $Q^2 \approx m_\mu^2$, through an interpolation.

The derivative of the VPF with respect to $Q^2$ is free of ultraviolet
divergences. The Adler function~\cite{Adler:1974gd,De Rujula:1976au}
is a related physical quantity, defined as follows,
\begin{equation}
  D(Q^2)=12\,\pi^2\,Q^2\, \frac{d\Pi(Q^2)}{dQ^2}\,.
  \label{eq:adler}
\end{equation}
The Adler function can be combined with an appropriate kernel function
to derive an alternative representation for $a_\mu^{\rm
  HLO}$~\cite{Jegerlehner:2008zza,Lautrup:1971jf},
\begin{equation}
  a_\mu^{\rm HLO} = \frac{\alpha^2}{6\pi^2} \int_{0}^{1}
  dx~ \frac{(1-x)(2-x)}{x} \,D\left(\frac{x^2m_\mu^2}{1-x}\right)\,,
  \label{eq:amuadler}
\end{equation}
where the substitution $Q^2 \to x^2m_\mu^2/(1-x)$ was applied. In this
way, lattice determinations of
$D(Q^2)$~\cite{Renner:2012fa,Francis:2013fzp,Horch:2013lla} can be
used to directly compute $a_\mu^{\rm HLO}$~\cite{DellaMorte:2014rta}.

The idea of taking the derivative of $\Pi(Q^2)$ with respect to $Q^2$
can be extended to include higher order derivatives at $Q^2=0$,
computed via Euclidean-time moments of the vector correlation
function, eq.~(\ref{eq:veccor}), at vanishing spatial
momentum~\cite{Chakraborty:2014mwa}. The subtracted VPF can then be
constructed from its Taylor expansion. Long-distance effects are
enhanced when increasing the order of the moments. For the $u,d$
contribution, these effects are expected to be sizeable since they are
related to the two-pion decay channel of the $\rho$-meson.

A new integral representation for $a_\mu^{\rm HLO}$ based on the
Mellin transform of the hadronic spectral
function~\cite{deRafael:2014gxa} relies on the calculation of the
moments ${\cal M}(-n)$,
\begin{equation}
  \label{eq:momeucl} {\cal M}(-n)= \frac{(-1)^{n+1} }{(n+1)!}(
  m_{\mu}^2 )^{n+1} \left.\frac{d^{n+1}}{(dQ^2
    )^{n+1}}\widehat{\Pi}(Q^2)\right|_{Q^2 =0}\,,
\end{equation}
with $n=\{0,1,2, \dots\}.$ In this approach, the subtracted VPF also
appears in the evaluation of integrals over $Q^2$, which are, however,
better suited than e.g. eq.~(\ref{eq:amulat}) for the regime of
momenta accessible on the lattice. An evaluation based on a
phenomenological model~\cite{Bernecker:2011gh} indicates that already
for the order $n=3$, a $1\%$ deviation from a determination of
$a_\mu^{\rm HLO}$ based on the dispersion relation approach could be
achieved~\cite{deRafael:2014gxa}.

\paragraph{Mixed (Time-Momentum) Representation}

Different representations for $a_\mu^{\rm HLO}$ can provide
alternative means to monitor the leading systematic effects present in
lattice computations -- a few examples have been mentioned above.
These integral representations can differ by the weight given to the
integrand by a particular $Q^2$ region or by the relative size of the
long-distance contributions. A representation could thus be better
suited for lattice QCD studies provided that it is more constrained by
the region where data is available and sufficiently accurate.

A {\it mixed-representation} of the subtracted VPF involving the
time-momentum dependence of the vector correlation function
$G(x_0,\,\vec k)$,
\begin{equation}
  {G(x_0,\,\vec k)} =
  {\int d^3 x\;} e^{i\vec k \vec x} \, \langle\, J_\mu(x_0,\vec x)\,
  J_\mu(0)\, \rangle \,,
  \label{eq:veccor}
\end{equation}
can be written as follows~\cite{Bernecker:2011gh},
\begin{equation}
  \widehat{\Pi}(Q^2) = \int_{0}^\infty dx_0\, {G(x_0,\vec k=0)}
  \left[x_0^2 - \frac{4}{Q^2}\sin^2\left(\frac{1}{2} Q x_0\right)
    \right]\,.
  \label{eq:mixrep}
\end{equation}
The subtracted VPF determined in this way preserves a continuous
dependence on $Q^2$, in particular in the neighbourhood of
$Q^2=0$~\cite{Feng:2013xsa,Francis:2013fzp,Francis:2014qta}.

The integration bounds in eq.~(\ref{eq:mixrep}) imply that
long-distance effects in $G(x_0,\, \vec k=0)$ will contribute. For
$u,d$ quarks, they are governed by the resonance nature of the
$\rho$-meson. This necessitates the incorporation of interpolating
operators which couple efficiently to two-pion states into the vector
correlation function. An appealing feature of the mixed-representation
is that quark-disconnected diagrams, which arise from the singlet
contribution to the vector correlation function, can be evaluated
straightforwardly using efficient noise reduction
techniques~\cite{Francis:2014hoa}.

Since different representations can lead to an improved control of the
uncertainties in distinct $Q^2$ intervals, it is beneficial to combine
the use of these representations to reduce the overall error on
$a_\mu^{\rm HLO}$. In general, a mixture of methods based on
previously discussed ideas -- used in combination with variance
reduction techniques~\cite{Blum:2012uh,Shintani:lat14} -- is expected
to lead to a more accurate lattice result for $a_\mu^{\rm HLO}$.
\begin{table}[t!]
  \begin{center}
    \begin{tabular}{cccccr}
      \hline
      Ens. & $a\,[\mathrm{fm}]$ & $V/a^4$ & $M_\pi$ & $M_\pi L$
      & $N_{\rm meas}$\\
      \hline
      $\sf A3$ & $0.079$ & $64 \times 32^3$  & $473$ & $6.0$ & $1004$\\
      $\sf A4$ &         & $64 \times 32^3$  & $363$ & $4.7$ & $1600$\\
      $\sf A5$ &         & $64 \times 32^3$  & $312$ & $4.0$ & $1004$\\
      $\sf B6$ &         & $96 \times 48^3$  & $267$ & $5.1$ & $1224$\\
      \hline
      $\sf E5$ & $0.063$ & $64 \times 32^3$  & $456$ & $4.7$ & $4000$\\
      $\sf F6$ &         & $96 \times 48^3$  & $325$ & $5.0$ & $1200$\\
      $\sf F7$ &         & $96 \times 48^3$  & $277$ & $4.2$ & $1000$\\
      $\sf G8$ &         & $128 \times 64^3$ & $193$ & $4.0$ & $820$\\
      \hline
      $\sf N5$ & $0.050$ & $96 \times 48^3$  & $430$ & $5.2$ & $1392$\\
      $\sf N6$ &         & $96 \times 48^3$  & $340$ & $4.1$ & $2236$\\
      $\sf O7$ &         & $128 \times 64^3$ & $261$ & $4.4$ & $552$\\
      \hline
    \end{tabular}
  \end{center}
  \caption{Ensembles of O$(a)$ improved Wilson fermions used in the
    determination of $a_\mu^{\rm HLO}$ by the Mainz group. Approximate
    values of the lattice spacing $a$ and of the pion mass $M_\pi$ (in
    $\mathrm{MeV}$) together with information about the lattice volume
    and the number of measurements $N_{\rm meas}$ are given.}
  \label{tab:ens}
\end{table}
%%

%%%%%%%%%%%%%%%%%%%%%%%%%%%%%%%%%%%%%%%%%%%%%%%%%%%%%%%%%%%%%%%%%%%

\section{Reaching the Physical Point}
\label{sec:physpoint}

We already mentioned that various sea and valence quark flavours
contributing to $a_\mu^{\rm HLO}$ are now being incorporated in the
lattice simulations (see Fig.~\ref{fig:gm2comp}). In addition,
simulations with non-degenerate $u$ and $d$ quark
masses~\cite{Gregory:2013taa} or studies of the valence $b$-quark
contribution with NRQCD~\cite{Colquhoun:2014ica} are also being
considered.

The approach to the physical point in the $u$,\,$d$ sector can be a
source of sizeable systematic effects. The light-quark mass dependence
of $a_\mu^{\rm HLO}$ is linked to the resonance nature of the
$\rho$-meson and is thus expected to become more important when
approaching the chiral limit. Different fit forms, often inspired by
chiral effective theories, have been used to estimate the uncertainty
from the chiral extrapolation. The explicit measurement of the vector
meson mass has also been used in the calculation of $a_\mu^{\rm HLO}$
to modify its chiral behaviour~\cite{Feng:2011zk}. Studies including
simulations in the neighbourhood of the physical point have recently
been
reported~\cite{Burger:2013jva,Gregory:2013taa,Chakraborty:2014mwa}. For
sufficiently large volumes, the physical effect of the $\rho$-meson
decay will contribute and a dedicated effort will be needed to address
the associated fluctuations.
\begin{figure}[t!]
  \centering
  \includegraphics[scale=0.69]{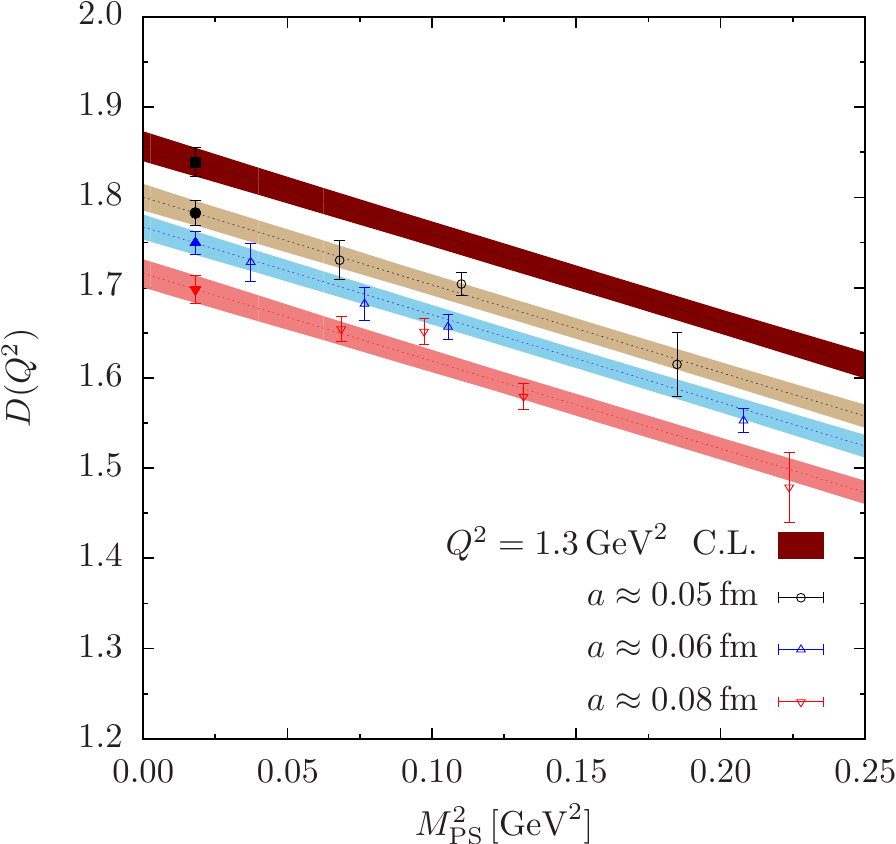}
  \caption{Pion-mass dependence of the Adler function $D(Q^2)$
    at fixed $Q^2=1.3\,{\rm GeV^2}$. The upper band, denoted by
    `C.L.' in the legend, is the continuum limit estimate. The
    leftmost (filled) symbols refer to the extrapolated values at the
    physical pion mass.}
  \label{fig:adler_mps}
\end{figure}
%% 

%%%%%%%%%%%%%%%%%%%%%%%%%%%%%%%%%%%%%%%%%%%%%%%%%%%%%%%%%%%%%%%%%%%

\section{Studies of $a_\mu^{\rm HLO}$ with improved Wilson fermions}
\label{sec:amuwilson}

The lattice group in Mainz has developed a dedicated research program
aiming at a precise determination of physical observables related to
the
VPF~\cite{DellaMorte:2011aa,Horch:2013lla,Francis:2013fzp,DellaMorte:2014rta,Francis:2014qta,Francis:2014hoa,Shintani:lat14,Herdoiza:2014jta,Francis:2014yga,DellaMorte:2012cf}. We
report some recent developments in the study of $a_\mu^{\rm HLO}$
where several of the previously discussed advances have been
implemented.

The lattice QCD ensembles (c.f. table~\ref{tab:ens}) with two
dynamical flavours of non-perturbatively O$(a)$ improved Wilson
fermions were produced as part of the CLS initiative. They include
three values of the lattice spacing $a$, large volumes and pion masses
down to $M_\pi\approx 190\,{\rm MeV}$. A substantial increase in the
number of measurements $N_{\rm meas}$ has also been achieved recently.

\begin{figure}[t!]
  \centering
  \includegraphics[scale=0.69]{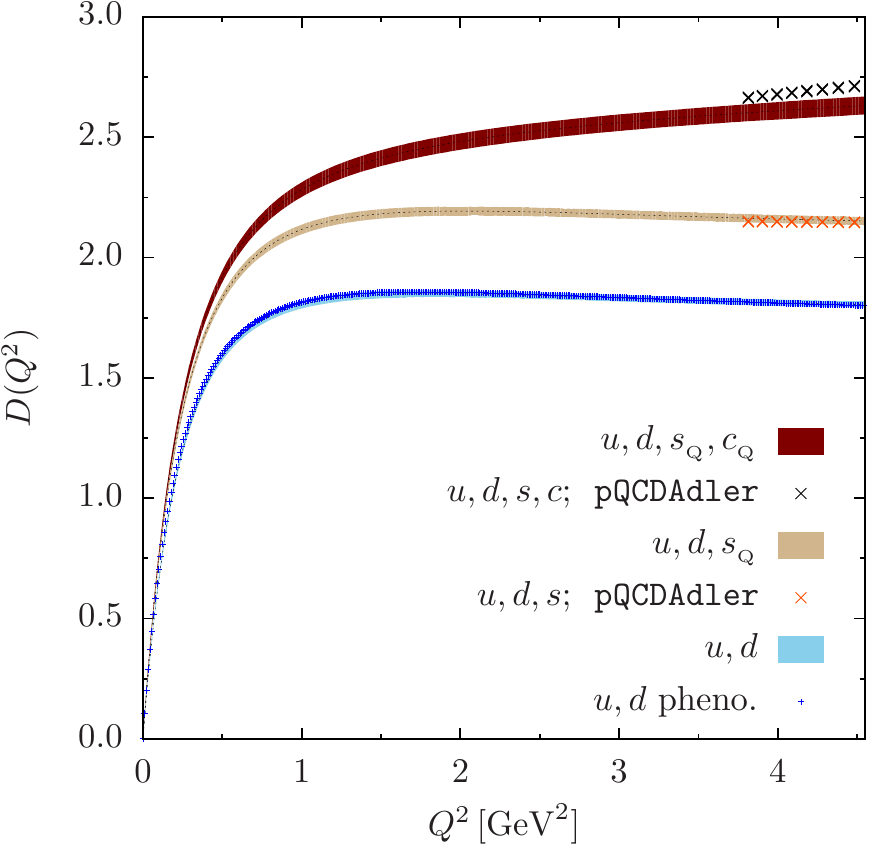}
  \caption{Contributions from $(u,d)$ and from partially quenched
    strange $s_Q$ and charm $c_Q$ quark flavours to the Adler function
    after having performed the continuum and chiral
    extrapolations. The $(u,d)$ contribution shows a good agreement
    with the phenomenological model of ref.~\cite{Bernecker:2011gh}
    denoted by the blue `+' symbols. For the cases where $s_Q$ and
    $c_Q$ are included, a comparison to perturbative QCD results from
    the \texttt{pQCDAdler} package~\cite{pqcdadler} is shown.}
  \label{fig:adler_fla}
\end{figure}
\begin{figure}[ht!]
  \centering
  \includegraphics[width=0.40\textwidth]{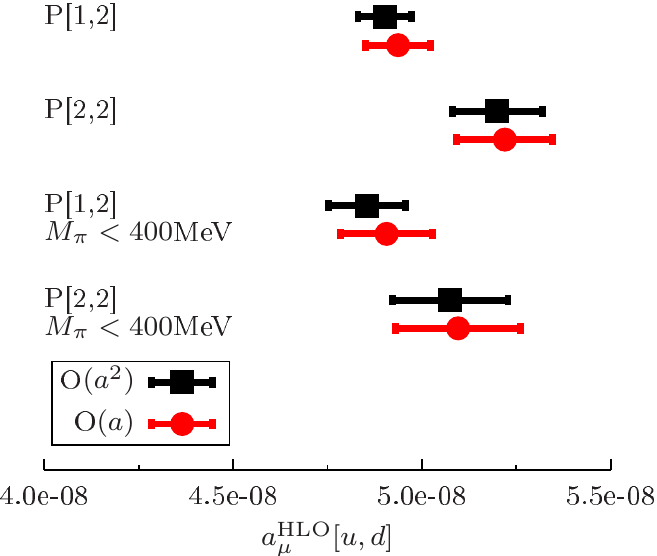}
  \caption{Preliminary results for the $u,d$ contribution to
    $a_\mu^{\rm HLO}$ based on the determination of the Adler function
    and on the use of the representation in
    eq.~(\ref{eq:amuadler}). The $Q^2$ dependence is examined by the
    use of Pad\'e approximants of order $[1,2]$ and $[2,2]$. Fits
    where pion masses, $M_\pi \geq 400\,{\rm MeV}$, have been
    included/excluded are used to study systematic effects in the
    light-quark mass dependence of $a_\mu^{\rm HLO}$. We observe that
    lattice artefacts are under control by performing separate
    analyses with fit ans\"atze including either O$(a)$ or O$(a^2)$
    terms~\cite{DellaMorte:2014rta}.}
  \label{fig:amu_adler}
\end{figure}
The VPF is extracted through eq.~(\ref{eq:vpcont}) from a lattice
determination of the vacuum polarisation tensor, using local-conserved
vector currents in the r.h.s of eq.~(\ref{eq:pol_tensor}). A high
density of $Q^2$ points for the VPF is attained by the use of
partially twisted boundary conditions. We take advantage of this in
order to derive the Adler function $D(Q^2)$ in eq.~(\ref{eq:adler})
from numerical derivatives of the
VPF~\cite{Horch:2013lla,DellaMorte:2014rta}. The $Q^2$ dependence of the
Adler function is then analysed in terms of Pad\'e approximants of
various orders. This study is integrated into a global analysis of
$D(Q^2)$ combining all the ensembles listed in table~\ref{tab:ens}. An
estimate of $D(Q^2)$ in the continuum limit and at the physical point
can be obtained in this way. An illustration of the pion-mass
dependence of the non-singlet $(u,d)$ contribution to the Adler
function at fixed $Q^2$ is shown in
Fig.~\ref{fig:adler_mps}. Systematic effects due to lattice artefacts
and from the extrapolation of the light-quark mass to the physical
point are explored by considering various fit forms and by repeating
the analysis on subsets of the available
ensembles~\cite{DellaMorte:2014rta}. The light $(u,d)$ as well as the
partially quenched strange $s_Q$ and charm $c_Q$ contributions to
$D(Q^2)$ are displayed in Fig.~\ref{fig:adler_fla}.  Some interesting
applications of the Adler function include the matching to
perturbation theory to determine the QCD coupling constant $\alpha_s$
or the study of the hadronic contribution to the running of QED
coupling~\cite{Herdoiza:2014jta,Francis:2014yga}. Preliminary results
for $a_\mu^{\rm HLO}$ from the use of the Adler function
representation in eq.~(\ref{eq:amuadler}) are shown in
Fig.~\ref{fig:amu_adler}.

\begin{figure}[t!]
  \centering \includegraphics[width=0.40\textwidth]{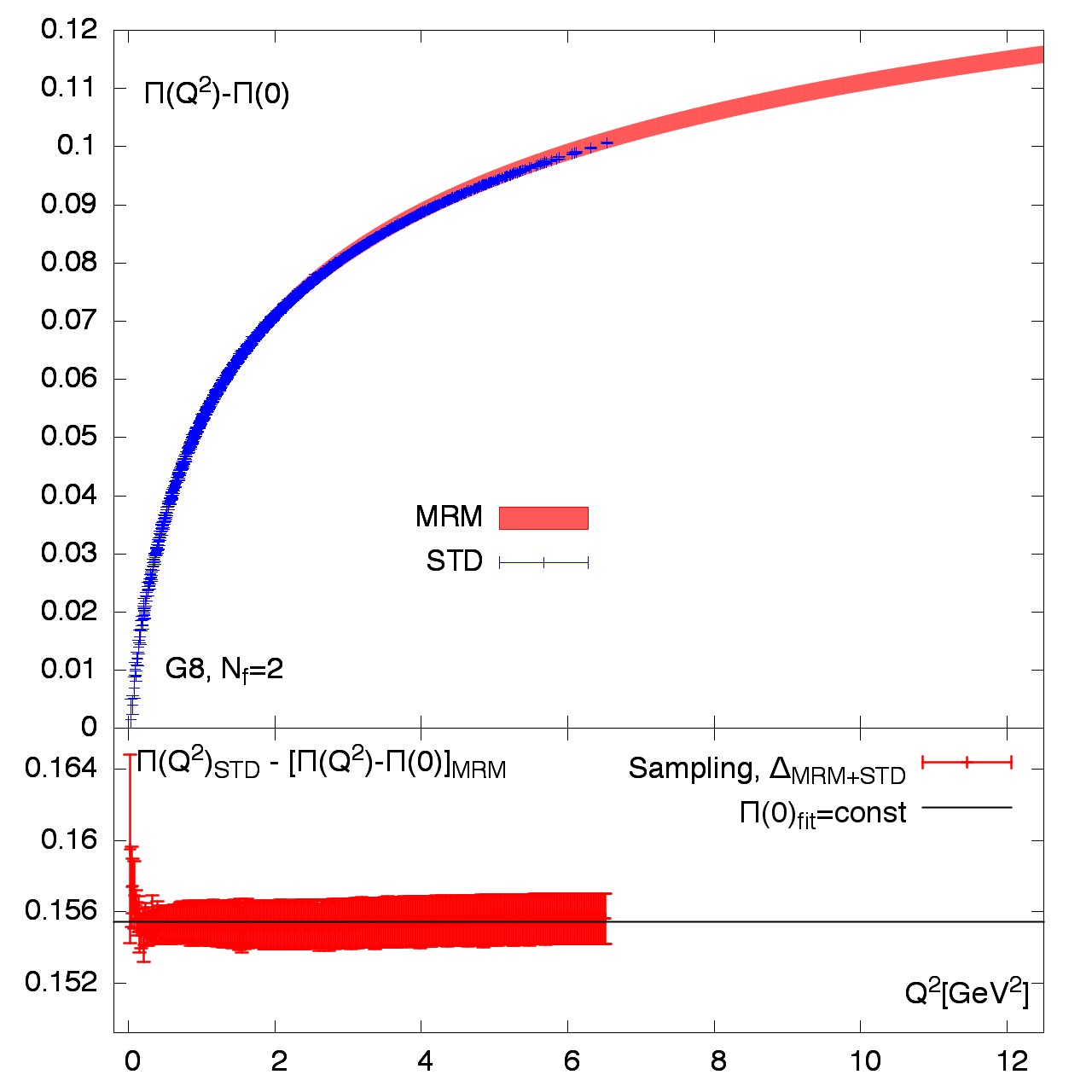}
  \caption{Comparison of the determinations of the subtracted VPF from
    the mixed-representation method (MRM) in eq.~(\ref{eq:mixrep}) and
    from the more standard approach (STD) based in
    eq.~(\ref{eq:vpcont}) and an extrapolation of $\Pi(Q^2)$ to
    $Q^2=0$. The upper panel shows the consistency among these methods
    for $\widehat{\Pi}(Q^2)$ over a large $Q^2$ interval. The lower
    panel shows the corresponding difference, $\Pi(Q^2)_{\rm STD}
    -\widehat{\Pi}(Q^2)_{\rm MRM}$ and demonstrates the stability of
    the derived values of $\Pi(0)$. Data from an ensemble with $M_\pi
    \approx 190\,{\rm MeV}$ are shown but similar results are observed
    for heavier pion masses up to $\sim 450\,{\rm MeV}$~\cite{Francis:2014qta}.}
  \label{fig:mixrep}
\end{figure}

The determination of the subtracted VPF from the {\it
  mixed-representation} in eq.~(\ref{eq:mixrep}) can be compared to
the more standard procedure where $\Pi(Q^2)$ is extracted from
eq.~(\ref{eq:vpcont}) and then extrapolated to $Q^2=0$ to determine
$\widehat{\Pi}(Q^2)$~\cite{Francis:2014qta}. The top panel of
Fig.~\ref{fig:mixrep} shows an example of this comparison for the
subtracted VPF over a large $Q^2$ interval. The agreement is
corroborated by the lower panel of Fig.~\ref{fig:mixrep} where the
extrapolated estimate for $\Pi(0)$ is checked against the
mixed-representation method.

\begin{figure}[t!]
  \centering
  \includegraphics[width=0.45\textwidth]{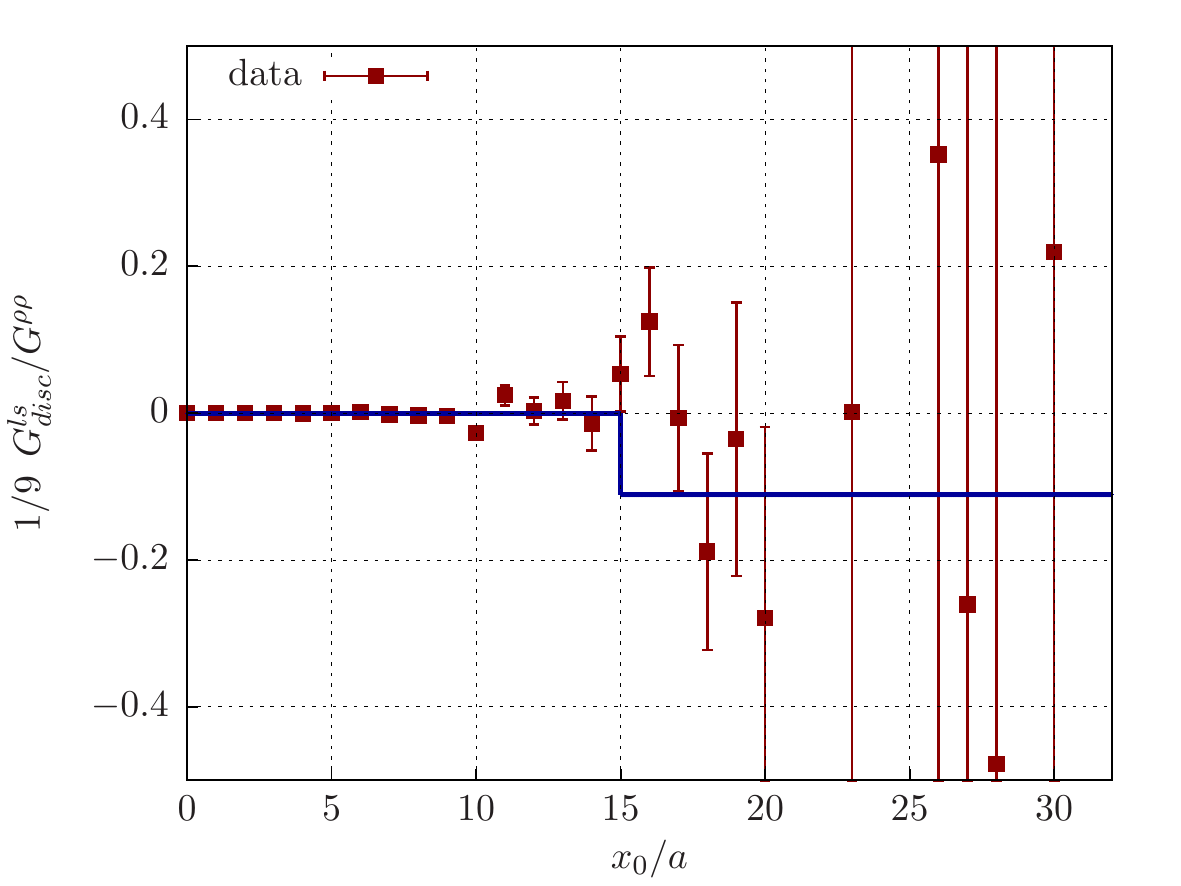}
  \caption{Lattice evaluation of the Euclidean time dependence of the
    ratio of the quark-disconnected vector correlation $G_{\rm
      disc}^{\ell s}(x_0)$, involving light $\ell=u,d$ and strange $s$
    quarks, to the isovector $\rho$-meson correlation function
    $G^{\rho\rho}(x_0)$~\cite{Francis:2014hoa}. The asymptotic value
    $-1/9$ in eq.~(\ref{eq:discon}) is denoted by the blue horizontal
    line for $x_0/a \geq 15$. Approximately $4 \cdot 10^{5}$
    inversions of the Dirac operator are needed to achieve the
    accurary shown in this figure.}
  \label{fig:discon}
\end{figure}

The flavour singlet currents in eq.~(\ref{eq:current}) require the
presence of Wick contractions involving both quark-connected and
quark-disconnected contributions to the vector correlation
functions. The latter suffer from large statistical fluctuations and
are therefore often neglected in present lattice computations due to
their high computational cost. It is however crucial to put a bound on
their expected size. The {\it mixed-representation} correlator
$G(x_0,\,\vec k=0)$ in eq.~(\ref{eq:veccor}) is dominated in the large
Euclidean time limit by the lowest energy state corresponding to the
isovector channel, i.e. $G^{\rho\rho}(x_0)$. This leads to the
following asymptotic behaviour~\cite{Francis:2013fzp,Francis:2014hoa}
of the quark-disconnected vector correlation $G_{\rm disc}^{\ell
  s}(x_0)$ involving light $\ell=u,d$ and strange $s$ quarks,
\begin{equation}
  \frac{1}{9}\,\frac{G_{\rm disc}^{\ell s}(x_0)}{G^{\rho\rho}(x_0)}
  \stackrel{x_0\to\infty}{\longrightarrow} -\frac{1}{9}\,,
  \label{eq:discon}
\end{equation}
in agreement with the expectation based on
ChPT~\cite{DellaMorte:2010aq}. A lattice evaluation of the l.h.s. of
eq.~(\ref{eq:discon}) as a function of $x_0$ is shown in
Fig.~\ref{fig:discon}. A significant reduction of the statistical
fluctuations in $G_{\rm disc}^{\ell s}(x_0)$ was obtained by using the
same stochastic sources for the light and strange quark
contributions. The signal is compatible with zero with an error
approaching $1/9$ at $x_0 \approx 15a \approx 1\,{\rm fm}$.  By
assuming that the asymptotic value in eq. (12) is reached at this
distance, a conservative upper bound on the disconnected contribution
of $\sim 4\%$ can be inferred.
%% 

%%%%%%%%%%%%%%%%%%%%%%%%%%%%%%%%%%%%%%%%%%%%%%%%%%%%%%%%%%%%%%%%%%%

\section*{Conclusions}

In the next few years, a new generation of experiments is expected to
improve the determination of the anomalous magnetic moment of the muon
$a_\mu$ by a factor of four. A similar improvement in the SM
prediction would greatly enhance the sensitivity to physics beyond the
SM. Leading order hadronic effects are responsible for the largest
theoretical uncertainty in $a_\mu$, coming from a phenomenological
approach based on a combination of dispersive techniques and
experimental data. Lattice QCD provides a first principles
determination that can lead to an independent and valuable check. We
have presented some recent ideas and applications that are expected to
lead to an improved determination of $a_\mu^{\rm HLO}$. Higher-order
hadronic effects from light-by-light scattering are the second largest
source of error in the SM prediction of $a_\mu$. We refer to
ref.~\cite{Shintani:ichep14} for a review presented at this conference
about the recent progress in using lattice QCD to determine these
contributions.

%%%%%%%%%%%%%%%%%%%%%%%%%%%%%%%%%%%%%%%%%%%%%%%%%%%%%%%%%%%%%%%%%%%

\section*{Acknowledgements}

We thank Michele Della Morte, Andreas J\"{u}ttner and Andreas Nyffeler
for useful discussions.
Our calculations were performed on the ``Wilson'' and ``Clover'' HPC
Clusters at the Institute of Nuclear Physics, University of Mainz.
We thank Dalibor Djukanovic and Christian Seiwerth for technical support.
This work was granted access to the HPC resources of the Gauss Center
for Supercomputing at Forschungzentrum J\"ulich, Germany, made available
within the Distributed European Computing Initiative by the PRACE-2IP,
receiving funding from the European Community's Seventh Framework
Programme (FP7/2007-2013) under grant agreement RI-283493 (project
PRA039).
We are grateful for computer time allocated to project HMZ21 on the
BG/Q JUQUEEN computer at NIC, J\"ulich.
This research has been supported in part by the DFG in the
SFB~1044.
We thank our colleagues from the CLS initiative for sharing
the ensembles used in this work.
G.H. acknowledges support by the Spanish MINECO through the Ram\'on y
Cajal Programme and through the project FPA2012-31686 and by the
Centro de Excelencia Severo Ochoa Program SEV-2012-0249.
%%



%% References

%% \nocite{*}
%% \bibliographystyle{elsarticle-num}
%% \bibliography{martin}






% \end{thebibliography}

%%%%%%%%%%%%%%%%%%%%%%%%%%%%%%%%%%%%%%%%%%%%%%%%%%%%%%%%%%%%%%%%%%%

\end{document}